\documentclass[final]{elsart3p}

\usepackage[dvips]{color,graphics}

\usepackage{graphicx}

 \journal{Journal of Alloys and Compounds}
\usepackage{dcolumn}
\usepackage{bm}

\begin{document}

\begin{frontmatter}

\title{Ab-initio design of half-metallic fully-compensated
ferrimagnets:
the case of Cr$_2$MnZ (Z= P, As, Sb, Bi) compounds}

\author[Patras]{I. Galanakis\corauthref{cor}}\ead{galanakis@upatras.gr}
\author[Gebze]{K. \"Ozdo\~gan}\ead{kozdogan@gyte.edu.tr}
\author[Julich,Fatih]{E. \c Sa\c s\i o\~glu}\ead{e.sasioglu@fz-juelich.de}
\author[Gebze]{B. Akta\c s}

\address[Patras]{Department of Materials Science, School of Natural
  Sciences, University of Patras,  GR-26504 Patra, Greece}
\address[Gebze]{Department of Physics, Gebze Institute of Technology,
Gebze, 41400, Kocaeli, Turkey}
\address[Julich]{Institut f\"ur Festk\"orperforschung, Forschungszentrum
J\"ulich, D-52425 J\"ulich, Germany}
\address[Fatih]{Fatih University,
Physics Department, 34500, B\" uy\" uk\c cekmece,  \.{I}stanbul,
Turkey} \corauth[cor]{Corresponding author. Phone +30-2610-969925,
Fax +30-2610-969368}

\begin{abstract}
Electronic structure calculations from first-principles are
employed to design some new half-metallic fully-compensated
ferrimagnets (or as they are widely known half-metallic
antiferromagnets) susceptible of finding applications in
spintronics. Cr$_2$MnZ (Z= P, As, Sb, Bi) compounds have 24
valence electrons per unit cell and calculations show that their
total spin moment is approximately zero for a wide range of
lattice constants in agreement with the Slater-Pauling behavior
for ideal half-metals. Simultaneously, the spin magnetic moments
of Cr and Mn atoms are antiparallel and the compounds are
ferrimagnets. Mean-field approximation is employed to estimate
their Curie temperature, which exceeds room temperature for the
alloy with Sb. Our findings suggest that Cr$_2$MnSb is the
compound of choice for further experimental investigations.
Contrary to the alloys mentioned above half-metallic
antiferromagnetism is unstable in the case of the Cr$_2$FeZ (Z=
Si, Ge, Sn) alloys.
\end{abstract}

\end{frontmatter}


\section{Introduction}\label{sec1}

The last decade a new field has evolved in solid state physics
focusing on the design of novel materials for spintronic
applications \cite{Zutic}. First-principles electronic structure
calculations have played a key role since several alloys for such
applications have been initially predicted before their synthesis
and their integration in realistic devices. Among these materials
the most promising ones are the so-called half-metals; magnetic
materials which are normal metals for one spin-direction and
semiconductors for the other and thus electrons near the Fermi
level are of a unique spin-character \cite{Review1,Review2}. The
first predicted half-metal was the ferromagnetic Heusler alloys
NiMnSb \cite{deGroot}.

The research on half-metallic ferromagnetic Heusler alloys is
intense and several such alloys have been predicted
\cite{GalaHalf,GalaFull}. An interesting case is also the
half-metallic ferrimagnets; compounds where the different
transition metal atoms in the unit cell have antiparallel magnetic
spin moments. Examples of such alloys are FeMnSb
\cite{GalaHalf,deGroot2} and Mn$_2$VAl
 \cite{Weht,Kemal,Sasioglu,ErsoyTc}. All these half-metals exhibit the
 so-called Slater Pauling behavior and the total spin moment in the unit cell
 is given as a simple function of the number of valence electrons. In the case of full-Heusler
 alloys having the chemical formula X$_2$YZ this rule takes the
 form $M_t=Z_t-24$ \cite{GalaFull}. $M_t$ is the total spin moment in the unit
 cell in $\mu_B$ and $Z_t$ the total number of valence electrons.
The number "24" comes from the fact that there are exactly 12
occupied electronic states in the semiconducting band. Thus
magnetic alloys with exactly 24 valence electrons should have
exactly zero total spin moment. Such a compounds would be
perfectly suited for applications due to the negligible external
magnetic field which it would create. The 24-valence electrons
compounds are known as half-metallic antiferromagnets, but their
correct definition is "half-metallic (HM) fully-compensated
ferrimagnets" (FCF)\cite{Leuken} and this is the reason why the
characteristic temperature for  these alloys is the Curie
temperature and not the N\'eel temperature as in usual
antiferromagnets. In the class of HM-FCF alloys belong the
hypothetical Heusler MnCrSb \cite{Leuken} and Mn$_3$Ga
\cite{Felser2006} compounds which have not been yet synthesized.
Another root towards HM-FCF is the doping of diluted magnetic
semiconductors \cite{Akai} and inclusion of Co-defects in
Mn$_2$VAl and Mn$_2$VSi alloys; Co atoms substitute Mn ones having
antiparallel moments between them \cite{GalaQuat,Gala-RC2}.

In this communication we will study the appearance of stable
half-metallic FCF in the case of the Cr$_2$MnZ alloys, where Z is
P, As, Sb or Bi which all correspond to a total of 24 valence
electrons. Cr and Mn atoms in these alloys have antiparallel spin
moments and the compounds show a gap in the spin-up band (we have
chosen it so that Cr atoms have positive spin moments and Mn atoms
negative spin moments). These alloys keep the HM character for a
wide range of lattice constants and the Fermi level behaves upon
compression or expansion of the lattice similar to a rigid band
model. Contrary to these alloys the Cr$_2$FeZ alloys (Z= Si, Ge or
Sn) which have also 24 valence electrons are not suitable for
realistic applications since they show a region of low density of
states (DOS) for a very narrow range of lattice parameters and any
deviation from these values completely destroys the low-DOS
region. For the calculations we employed the full--potential
nonorthogonal local--orbital minimum--basis band structure scheme
(FPLO) within the local spin density approximation to the density
functional \cite{koepernik}. We should also note that we have used
the scalar relativistic formulation and thus the spin-orbit
coupling was not taken into account. This is expected to play a
crucial role only in the case of the compound containing Bi
\cite{Mavropoulos} where the image of the spin-down states in the
spin-up band is not negligible as for the other lighter chemical
elements.

Finally we apply the augmented spherical wave (ASW) method
\cite{asw} in conjunction with the  frozen-magnon approach
\cite{magnon} to calculate the interatomic exchange interactions
which are used to estimate the Curie temperature of the more
technological relevant Cr$_2$MnZ compounds within the
multi-sublattice mean field approximation already employed in the
case of other Heusler alloys (see Refs.
\cite{ErsoyTc,Kubler,HeuslersTc}).

\begin{figure}
\begin{center}
\includegraphics[scale=0.55]{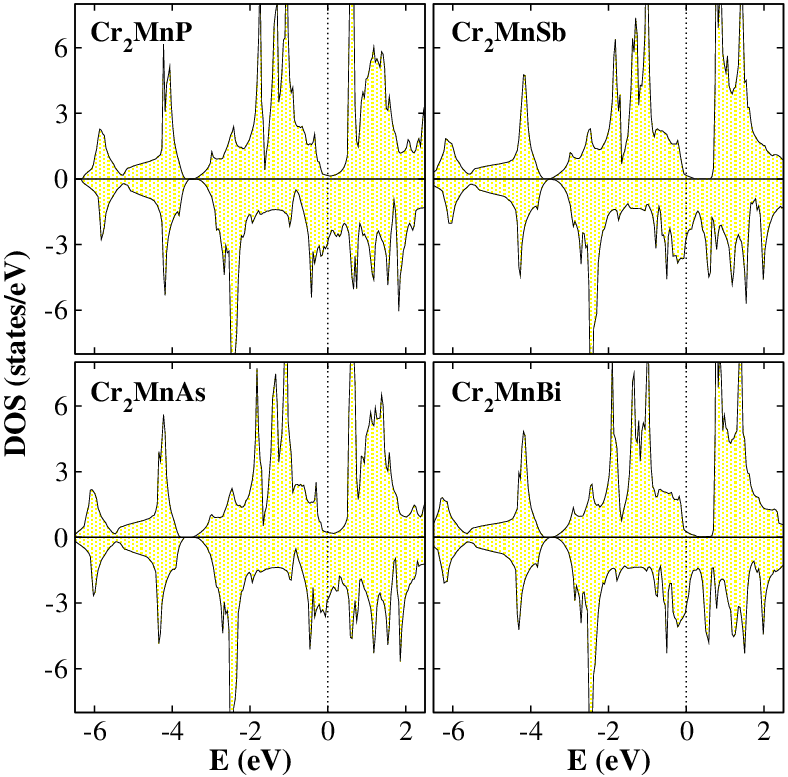}
\end{center} \caption{(Color online) Total density of states (DOS) for the four compounds under
study and for a lattice constant of 6.2 \AA . All four alloys are
almost half-metals. Positive values of DOS correspond to spin-up
electrons and negative values to spin-down electrons. The energy
scale is such that the zero energy corresponds to the energy of
the Fermi level. We have chosen to present the gap in the spin-up
band so that Cr atoms have positive spin moments and Mn atoms
negative spin-moments. \label{fig1}}
\end{figure}

\section{Cr$_2$MnZ (Z=P,As,Sb,Bi) alloys}\label{sec2}

We will start our discussion from the case of Cr$_2$MnZ alloys.
Since we want our compounds to have exactly 24 electrons, we
choose Z to be one of the isovalent P, As, Sb or Bi. Usually, in
the case of the full-Heusler alloys of the chemical type X$_2$YZ,
X atoms have higher valence than the Y ones, e.g. the X atom has
more valence electrons. In the compounds which we study, the
opposite occurs. Moreover, state-of-the-art methods for synthesis
of thin films like the Molecular Beam Epitaxy make possible to
grow novel materials as multilayers or thin-films where the spacer
or the substrate are responsible for the lattice constant adopted
by the material which we want to grow. Thus the Heusler alloys of
the type Cr$_2$MnZ  presented in this study could be eventually
grown experimentally with their lattice constant determined by an
adjutant in a way that the desired electronic and magnetic
properties appear.
\begin{figure}
\begin{center}
\includegraphics[scale=0.55]{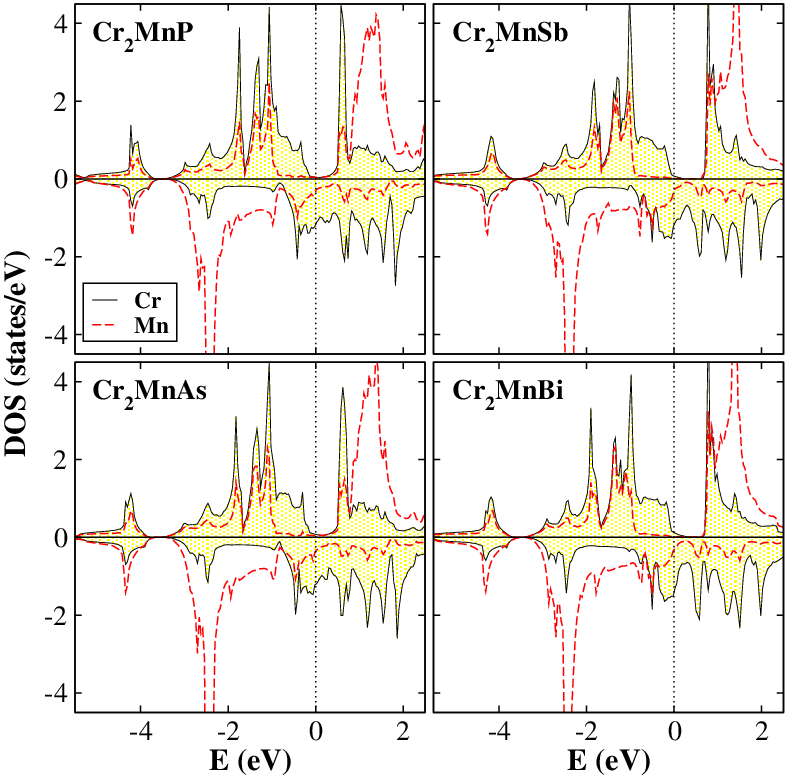}
\end{center} \caption{(Color online) Cr and Mn resolved DOS for the four case presented
in Fig. \ref{fig1}. Cr-DOS has been scaled to one atom.
\label{fig2}}
\end{figure}
For this reason we have studied their properties as a function of
the lattice constant. In Fig. \ref{fig1} we present the total
density of state (DOS) per unit cell for all four compounds and
for a lattice constant a=6.2\AA . We have chosen the spin-up
states such that Cr atoms have positive spin moments and Mn
negative ones. For all four compounds under study and for this
specific lattice constant there is a real gap in the spin-up
states for the alloys containing the heavier Sb and Bi atoms while
there is a region of tiny spin-up DOS for the compounds with P and
As (Note that since there is equal number of states for both
spin-directions we do not employ the terms majority and minority
to describe the spin-bands). Since the spin-polarization at this
region for the two latter compounds is almost 100\% we can safely
state that all four compounds present a real gap, the width of
which is around 0.5-0.7 eV. In the case of the lighter P- and
As-based compounds the Fermi level falls within this gap and the
Cr$_2$MnP and Cr$_2$MnAs are true half-metals while for the Sb-
and Bi-based alloys the Fermi level is just below the left edge of
the gap but the spin-polarization is still almost 100\% ,
partially thanks to the very large value of the spin-down DOS at
this region.

The gap is created due to the large exchange-splitting of both the
Cr and Mn atoms as can be seen in Fig. \ref{fig2} where we present
the atom-resolved DOS for the same lattice constant a=6.2\AA . We
do not present the DOS for the sp atoms since this is very small
with respect to the transition metal atoms. The occupied spin-up
states are mainly of Cr character while the occupied spin-down
states are mainly located at the Mn atoms. Thus the large exchange
splitting between the occupied states of one spin-direction and
the unoccupied states of the opposite spin direction for both Mn
and Cr atoms are added up to open the gap. We should also mention
that just below the gap in the spin-up band the states are made
exclusively of Cr states. This can be understood in terms of the
discussion in Ref. \cite{GalaFull} where it was shown in the case
of Co$_2$MnGe that the states just below the gap are the triple
degenerated $t_{1u}$ and just above the gap the double degenerated
$e_u$. Both $t_{1u}$ and $e_u$ were located exclusively at Co
sites and they did not hybridize with Mn states due to symmetry
reasons. In the compounds under study Cr plays the role of Co.
Just below the gap there are exclusively Cr $t_{1u}$ states, while
the larger exchange splitting of the Cr atoms with respect to the
Co ones pushes the Cr $e_u$ states higher in energy and they
cannot be distinguished from the other states at the same energy
region.

\begin{figure}
\begin{center}
\includegraphics[scale=0.55]{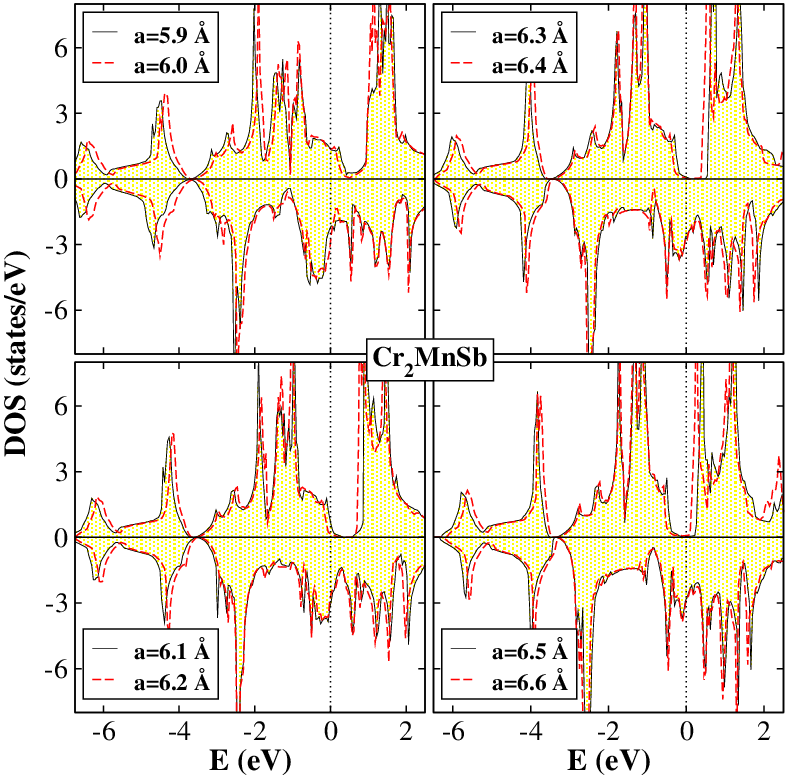}
\end{center} \caption{(Color online) Total DOS for the Cr$_2$MnSb alloy as a function of the
lattice constant. As the lattice is expanded the Fermi level is
shifted higher in energy similar to a rigid band model.
\label{fig3}}
\end{figure}

Our results up to now concern only one lattice constant the 6.2\AA
. The question which arises now is if these four alloys present
half-metallicity only for this specific lattice constant and how
the gap behaves with the lattice parameter. In Fig. \ref{fig3} we
present the total DOS for Cr$_2$MnSb for an ensemble of 8
different lattice constants ranging between 5.9 and 6.6 \AA\  as
an example since all four compounds present similar behavior . The
spin-up band presents only marginal changes with the lattice
constant and the gap persists for all the range of the lattice
parameters which we studied. What differs is the position of the
Fermi level, which behaves as in a rigid-band model. If we start
from the 6.2 \AA\ lattice parameter and we compress in a uniform
way the lattice, the Fermi level is shifted lower in energy with
respect to the gap, while the opposite occurs when we expand the
lattice where the Fermi level is shifted higher in energy.   Since
the number of valence electrons is fixed to 24, when the Fermi
level is out of the gap-bounds small changes occur also in the
spin-down band to take into account the extra charge in case that
the Fermi level is below the gap or the missing charge when the
Fermi level is above the gap. This behavior of the Fermi level is
due to the $p$ states of the sp atom and is opposite to the
behavior of the full-Heusler alloys like Co$_2$MnSi
\cite{Review2}. Thus for all four compounds we can determine the
range of lattice constants for which they are half-metallic. Our
findings show that the Cr$_2$MnP and Cr$_2$MnAs alloys are
half-metals between 6 and 6.2 \AA , while Cr$_2$MnSn and
Cr$_2$MnBi are half-metals between 6.2 and 6.6 \AA . Here we
should note that as mentioned in the introduction we have used the
scalar-relativistic approximation and we have not taken into
account the spin-orbit coupling. As previous calculations on
half-metals (see Ref. \cite{Mavropoulos}) have shown this is
important only for compounds containing Bi like Cr$_2$MnBi and
thus we expect the latter compound to be less likely to present
half-metallicity. Moreover, the alloys containing P and As are
probably difficult to be grown in such an expanded lattice
constant demanded by the presence of half-metallicity. Thus
Cr$_2$MnSb is probably the ideal case for experimentalists to try
to synthesize.

\begin{table}
\caption{Total and atom-resolved spin magnetic moments in $\mu_B$
for the Cr$_2$MnZ compounds where Z= P, As, Sb or Bi and for
different values of the lattice constant. The alloys containing P
or As are half-metals between 6.0 and 6.2 \AA , while the ones
containing Sb or Bi between 6.2 and 6.6 \AA . Note that the atomic
moments have been scaled to one atom. Last column are the
estimated Curie temperatures using the multi-sublattice mean-field
approximation \cite{Kubler,ErsoyTc}.}
 \begin{tabular}{lccccc} \hline \hline
  \multicolumn{6}{c}{Cr$_2$MnP} \\
 a(\AA ) & $m^{Cr}$  &  $m^{Mn}$  & $m^{P}$ & $m^{Total}$ & $T_C$ \\

6.0 & 1.528 & -3.044 & -0.071 & -0.060 &   \\

6.2& 1.796 & -3.392 & -0.098 & 0.102 &  240 \\

 \hline
 \multicolumn{6}{c}{Cr$_2$MnAs} \\
 a(\AA ) & $m^{Cr}$  &  $m^{Mn}$  & $m^{As}$ & $m^{Total}$ & $T_C$ \\

6.0 & 1.515 & -3.051 & -0.124 & -0.121 &   \\

6.2& 1.805 & -3.391 & -0.099 & 0.096 &  250 \\
 \hline

 \multicolumn{6}{c}{Cr$_2$MnSb} \\
 a(\AA ) & $m^{Cr}$  &  $m^{Mn}$  & $m^{Sb}$ & $m^{Total}$ & $T_C$ \\

6.2 & 1.670 & -3.277 & -0.100 & -0.036 &  342 \\

6.4 & 1.854 & -3.589 & -0.119 & -0.001 &  \\

6.6 & 2.029 & -3.874 & -0.141 & 0.043 & \\
 \hline
  \multicolumn{6}{c}{Cr$_2$MnBi} \\
 a(\AA ) & $m^{Cr}$  &  $m^{Mn}$  & $m^{Bi}$ & $m^{Total}$ & $T_C$ \\

6.2 & 1.608 & -3.224 & -0.096 & -0.011 &  320 \\

6.4 & 1.825 & -3.543 & -0.073 & -0.083 &  \\

6.6 & 1.931 & -3.687 & -0.108 & 0.068 & \\

\hline \hline
\end{tabular}
\label{table1}
\end{table}

We have not yet discussed the behavior of the magnetic spin
moments in these alloys. We have gathered in Table \ref{table1}
both the total and atom-resolved spin moments for all four
compounds and for the lattice constants where half-metallicity is
present. This is verified by the calculated total spin moments
which are very close to the ideal value of zero $\mu_B$. The Cr
and Mn spin moments are very close to the usual values which they
adopt in intermetallic compounds. Each Cr atom has a spin moment
around 1.5 to 2 $\mu_B$ while Mn have antiparallel spin moments
which are approximately two times the Cr ones in order to achieve
the zero total spin moment in the unit cell. The antiparallel
coupling of the moments is expected. Each Cr atom has four Mn (and
four sp atoms) as first neighbors and each Mn atom has eight Cr
atoms as nearest neighbors and as usually occurs for these
specific transition metal atoms, when they are very close in
space, they couple antiferromagnetically. As we expand the
lattice, we decrease the hybridization between neighboring atoms
and we increase their atomic-like character and thus enhance their
spin moments. The sp atoms have very small spin moments
antiparallel to the Cr ones occupying the X sites in the lattice.
This behavior is well known in most Heusler alloys (see Ref.
\cite{Review1} for a discussion on this coupling).

Besides high spin polarization of the states at the Fermi level,
an important further condition  for spintronics materials is a
high Curie temperature which should exceed the room temperature
for realistic applications. To estimate the Curie temperatures for
the half-metallic Cr$_2$MnZ compounds, we first applied the ASW
method in conjunction with the frozen-magnon approach to calculate
the interatomic exchange interactions \cite{asw,magnon}. Then the
calculated exchange parameters are used to estimate the Curie
temperature within multi-sublattice mean field approximation
\cite{HeuslersTc}. Note that we do not present the  exchange
parameters here in order to keep the discussion short. The
multi-sublattice  mean field approximation has already  been
employed to other  Heusler alloys and the results were in
reasonable agreement with the experimental data
\cite{ErsoyTc,Kubler,HeuslersTc}. Thus this approximation can be
used for trustworthy estimations. We have calculated the Curie
temperature for the lattice constant of 6.2 \AA\  but it should
vary only marginally for small changes of the lattice constants.
We present our results at the last column of Table \ref{table1}.
For the Cr$_2$MnP and Cr$_2$MnAs alloys the Curie temperature was
found to be 240-250 K and for Cr$_2$MnBi around 320 K which is not
suitable for realistic applications. For the Sb alloy it was found
to exceed the room temperature being 342 K close to the Curie
temperature of Co$_2$CrAl which is also under intense
investigation \cite{landolt}.

\section{Cr$_2$FeZ (Z=Si,Ge,Sn) alloys}\label{sec3}

\begin{figure}
\begin{center}
\includegraphics[scale=0.55]{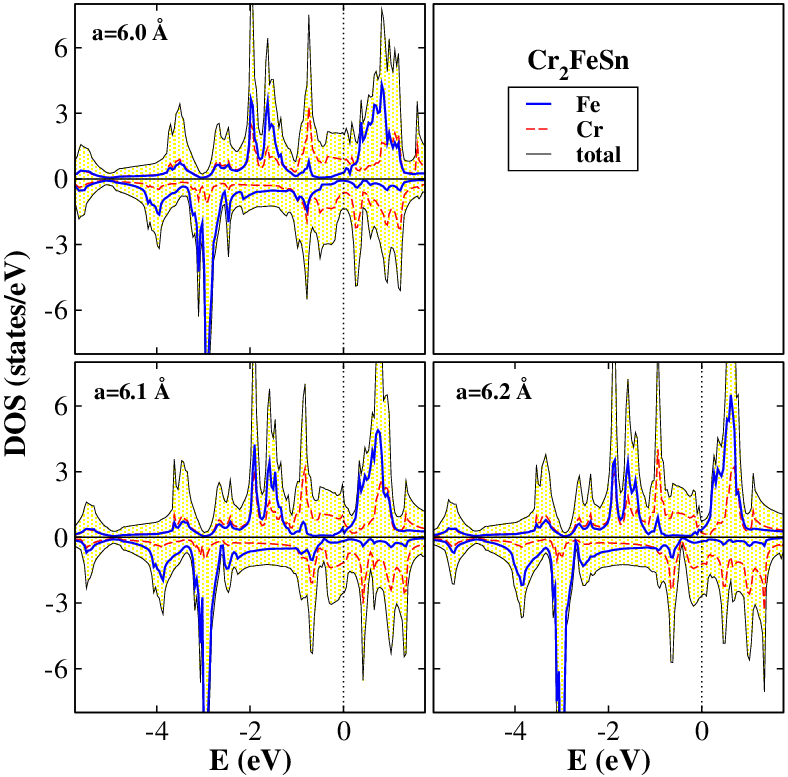}
\end{center} \caption{(Color online) Total, Cr- and Fe-resolved DOS for the
Cr$_2$FeSn alloy and for three different values of the lattice
constant. As the lattice is compressed, the region of low total
spin-up DOS  vanishes. Cr$_2$FeSi and Cr$_2$FeGe show similar DOS.
\label{fig4}}
\end{figure}

Motivated by our results on the Cr$_2$MnZ alloys, we decided to
study also another family of 24-valence electrons compounds
containing Fe instead of Mn: the Cr$_2$FeZ alloys where Z is Si,
Ge or Sn. In the right lower panel of Fig. \ref{fig4} we present
the total and the Cr and Fe-resolved DOS for a lattice constant of
6.2 \AA\ for the Cr$_2$FeSn alloy (the other two compounds
Cr$_2$FeSi and Cr$_2$FeGe present similar behavior). For this
lattice constant all three compounds under study present a region
of low DOS in the spin-up band and the Fermi level falls within
this region. Contrary to the compounds containing Mn, where the
spin-up occupied states are mainly of Cr character and the
occupied spin-down states are of mainly Mn character, for the
Fe-based alloys both the spin-up and spin-down occupied states
exhibit a more mixed character. The electronic structure is more
complicated than the Mn-based alloys and as a result when we
slightly vary the lattice constant the band-structure changes in a
way that the the region of low DOS is completely destroyed. This
is illustrated again in Fig. \ref{fig4} where we present also the
DOS for  Cr$_2$FeSn and for lattice constants of 6.0 and 6.1 \AA.
Cr spin-up states below the Fermi level and the Cr spin-up states
just above the Fermi level move one towards the other completely
destroying the region of high spin-polarization. The reason is the
different hybridization between the transition-metal atoms when we
substitute Fe for Mn. The smaller exchange splitting of the Fe
atoms closes the gap and this leads also to a different
distribution of the Cr charge which no more shows a large band-gap
as for the Mn-based alloys. To summarize our findings, the
calculations suggest that the region of low spin-up DOS persists
for Cr$_2$FeSn between 6.2 and 6.3 \AA\, and for Cr$_2$FeSi and
Cr$_2$FeGe between 6.1 and 6.2 \AA .

In Table \ref{table2} we present the spin-moments for the three
compounds and for the lattice constant of 6.2 \AA . Fe atoms can
not surpass the barrier of 3 $\mu_B$ since the extra electron with
respect to the Mn atom occupies mostly spin-up states reducing the
Fe spin-moment with respect to the Mn atoms. As a result also the
Cr spin moments are smaller. Overall the Cr$_2$FeZ alloys with
24-valence electrons are not suitable for realistic application,
contrary to their Cr$_2$MnZ compounds since half-metallicity is
very fragile in their case and thus we have not performed extra
calculations for their Curie temperatures.

\begin{table}
\caption{Total and atom-resolved spin magnetic moments in $\mu_B$
for the Cr$_2$MFeZ compounds where Z= Si, Ge,  or Sn and for a
lattice constant of 6.2 \AA . The Si and Ge have a total spin
moment very close to zero for lattice constants between 6.1 and
6.2 \AA\ and the Sn alloy between 6.2 and 6.3 \AA .}
 \begin{tabular}{lcccc} \hline \hline
Compound  & $m^{Cr}$  &  $m^{Fe}$  & $m^{Z}$ & $m^{Total}$ \\
 Cr$_2$FeSi  & 1.527 & -2.856 & -0.106 & 0.092 \\

 Cr$_2$FeGe & 1.535 & -2.853 & -0.140 & 0.077 \\

  Cr$_2$FeSn  & 1.447 & -2.789 & -0.111 & -0.007 \\

\hline \hline
\end{tabular}
\label{table2}
\end{table}

\section{Conclusion}\label{sec4}

We have employed ab-initio electronic structure calculations to
design some new half-metallic fully-compensated ferrimagnets (or
as they are widely known half-metallic antiferromagnets)
susceptible of finding applications in spintronics. Cr$_2$MnZ (Z=
P, As, Sb, Bi) compounds have 24 valence electrons per unit cell
and calculations show that their total spin moment is
approximately zero for a wide range of lattice constants in
agreement with the Slater-Pauling behavior for ideal half-metals.
Cr and Mn atoms have large antiparallel spin moments and the these
compounds are ferrimagnets. Upon expansion or compression of the
lattice, the Fermi level is shifted as in a rigid-band model.
Moreover mean-field approximation was employed to estimate their
Curie temperature, which exceeds room temperature for the alloy
with Sb. Cr$_2$MnSb has significant advantages with respect to the
other three compounds - larger Curie temperature, small influence
of spin-orbit coupling, large range of lattice constants where
half-metallicity is present - and we expect it to be the alloy of
choice for further experimental researches.

Contrary to these alloys, half-metallic antiferromagnetism is
unstable in the case of the Cr$_2$FeZ (Z= Si, Ge, Sn) alloys which
show a region of low density of states instead of a gap for a very
narrow range of lattice constants.

 Authors  acknowledge  the computer support of the ``Leibniz Institute for Solid State and
Materials Research Dresden''.


\end{document}